\documentclass[nofootinbib,preprint,aps,draft,pre,superscriptaddress,citeautoscript,floatfix]{revtex4-1}

\usepackage{graphics}
\usepackage{graphicx}
\usepackage{mathrsfs}

\begin{document}

\title{Ring to Mountain Transition in Deposition Pattern of Drying Droplets}

\author{Xingkun Man}
\affiliation{Center of Soft Matter Physics and its Applications, Beihang University, Beijing 100191, China}
\affiliation{School of Physics and Nuclear Energy Engineering, Beihang University, Beijing 100191, China}
\author{Masao Doi$^{*,}$}
\affiliation{Center of Soft Matter Physics and its Applications, Beihang University, Beijing 100191, China}
\affiliation{School of Physics and Nuclear Energy Engineering, Beihang University, Beijing 100191, China}

\begin{abstract}

When a droplet containing a non-volatile component is dried on a substrate,
it leaves a ring-like deposit on the substrate. We propose a theory which predicts the
deposit distribution based on a model of fluid flow and contact line motion of the droplet.
It is shown that the deposition pattern changes
continuously from coffee-ring to volcano-like and to mountain-like
depending on the mobility of the contact line and the evaporation rate.  Analytical
expression is given for the peak position of the distribution of deposit left on the substrate.
\end{abstract}

\maketitle

\begin{figure}[h]
\begin{center}
\includegraphics[bb=0 0 335 230, scale=0.75,draft=false]{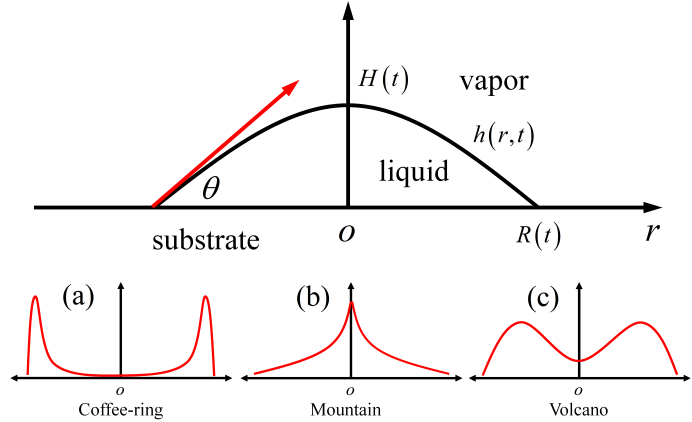}
\caption{(Color Online).
Schematic of a droplet and three different deposition patterns with axisymmetry in a cylindrical coordinate system (side-view). Relevant parameters are the radius of the contact line, $R(t)$, the height of the droplet at the center, $H(t)$, the contact angle, $\theta(t)$, and the profile of liquid/vapor interface, $h(r,t)$.}
\label{fig1}
\end{center}
\end{figure}

Drying of particle suspensions and polymer solutions gives rise to surprisingly rich deposition
patterns~\cite{Deegan00a,Bonn09}. The most well-known one is the coffee-ring pattern
of colloidal suspensions (Fig.\ref{fig1}(a)),
where all particles in the droplet are swept to the edge of the droplet.
This happens when the contact line is pinned~\cite{Deegan1997,Marin11}.
Mountain-like deposits (Fig.\ref{fig1}(b)), which have peaks at the droplet center, have also been
reported when the contact line are receding~\cite{Willmer2010,Li2013,Li2014}.
Multi-ring patterns have been observed when contact lines show stick-slip
motion~\cite{Lin2005,Xu2006,Mahe2008}, and  volcano-like patterns (Fig.\ref{fig1}(c))
have also been reported in drying of polymer solutions~\cite{Kajiya2009,Fukuda2013}.

Theoretical models have been developed for the drying phenomena of volatile droplets.
Previously, most works were limited to the case of ``pinned" contact line.
Deegan et al.~\cite{Deegan2000} first analysed the flow induced in liquid droplet
by evaporation and explained the physical origin of the coffee-ring.
Hu and Larson~\cite{Larson2002,Larson2005} carefully compared the results of flow simulation with
experiments and pointed out Marangoni effect is important. Such works
have further been extended to calculate the profile of the deposit left on the
substrate~\cite{Parisse97, Kobayashi, Okuzono}.

Compared with the case of pinned contact line, much less works have been done for
the case of moving contact line. Frastia et al.~\cite{Frastia2011} have conducted simulation assuming  a concentration
dependent viscosity, and have shown that multi-ring deposition patterns are obtained.
Freed-Brown~\cite{Brown2014} made a simple theory assuming that contact line
can move freely to keep the contact angle at the equilibrium value, and explained the
mountain-like deposition pattern. More recently, Kaplan et al.~\cite{Kaplan2015}
proposed a model to interpret the transition of deposition pattern from uniform films to single rings.
All these theories use different models for the motion of the contact line, and our understanding
for the deposition pattern in droplets having moving contact line is still at a primitive stage.

In this paper, we shall propose a simple model for the drying droplet with
moving contact line, and discuss the change of the deposition pattern (ring to mountain) when the mobility of the contact line and evaporation rate are changed.

We consider a liquid droplet placed on a substrate.  Let $R(t)$ be the radius of the base
circle of the droplet, and $H(t)$ be the height at the center.
We assume that the contact angle is small ($R(t) \gg H(t)$) and therefore the height
profile of the droplet $h(r,t)$ at time $t$ is given by a parabolic function
\begin{equation}\label{f1}
  h(r,t)=H(t)\left[1-\frac{r^2}{R^2(t)}\right].
\end{equation}
The contact angle is given by $\theta(t)= 2H(t)/R(t)$, and  the droplet volume $V(t)$ is
\begin{equation}\label{f2}
  V(t)=\frac{\pi}{2}H(t)R^2(t).
\end{equation}

The volume $V(t)$ of the droplet decreases in time due to solvent evaporation.
The evaporation rate of a droplet is determined by the diffusion of solvent molecules
in gas phase, and can be analyzed theoretically.  It has been shown that  $\dot V(t)$ is
proportional to the base radius of the droplet \cite{Parisse97, Kobayashi}, and weakly dependent
on the contact angle $\theta$.  Here we ignore the contact angle dependence, and write
$\dot V(t)$  as
\begin{equation}\label{f3}
  \dot V(t)= \dot V_0 \frac{R(t)}{R_0}
\end{equation}
where $\dot V_0$ ($ <0$) and $R_0$ denote the initial values of $\dot V(t)$ and $R(t)$ respectively.
$\dot V_0$ is expressed in terms of the diffusion constant of solvent molecules in gas phase,
the solvent vapor pressure near and far from the droplet, and the initial shape of the
droplet.

Given the equation for $\dot V(t)$, we need one more equation for  $\dot R(t)$.
To determine this, we use the Onsager  principle~\cite{Qian2006, Doi2011,Doi2013, Doi2015}.
In the current problem, this principle is equivalent to the minimum energy
dissipation principle in Stokesian hydrodynamics which states that the evolution of the
system is determined by the minimum of Rayleighian defined by
\begin{equation}\label{f4}
\Re=\Phi+\dot{F}
\end{equation}
where $\dot F$ is the time derivative of the free energy of the system, and
$\Phi$ is the energy dissipation function.

The free energy $F$ is easily obtained. The droplet size is assumed to be less than
the capillary length, then the free energy $F$ is written as a sum of the interfacial energy
\begin{eqnarray}\label{f5}
   F &=& (\gamma_{LS}-  \gamma_{SV}) \pi R^2 +
        \gamma_{LV} \int^R_0 dr 2\pi r \sqrt{ 1 + h'(r)^2 }
                                   \nonumber \\
    &=& (\gamma_{LS} -  \gamma_{SV}+ \gamma_{LV} )\pi R^2   + \gamma_{LV} \pi H^2
\end{eqnarray}
where $\gamma_{LV}$, $\gamma_{LS}$, and $\gamma_{SV}$, are the interfacial energy density
at liquid/vapor,  liquid/substrate and substrate/vapor interfaces, respectively.
Using the equilibrium contact angle
$\theta_e= [2(\gamma_{LV}+ \gamma_{LS}- \gamma_{SV})/\gamma_{LV} ]^{1/2}$,
and the volume $V$ of the droplet, eq.~(\ref{f5}) is written as
\begin{equation}\label{f6}
F=\gamma_{LV}\left(\frac{4V^2}{\pi R^4}+\frac{\pi R^2\theta^2_e}{2}\right).
\end{equation}
Then, the time derivative of such free energy is easily obtained as
\begin{equation}\label{f7}
\dot{F}=\gamma_{LV}
      \left[ \left( -\frac{16V^2}{\pi R^5}+\pi\theta^2_e R \right ) \dot R
                +\frac{8V\dot{V}}{\pi R^4}
      \right].
\end{equation}

To calculate the energy dissipation function $\Phi$, we use the lubrication approximation.
Let $v(r,t)$ be the height-averaged fluid velocity at position $r$ and time $t$. The
energy dissipation function $\Phi$ is written as
\begin{equation}\label{f8}
\Phi=\frac{1}{2}\int^{R}_{0}dr2\pi r\frac{3\eta}{h}v^2+\pi\xi_{\rm{cl}} R\dot{R}^2.
\end{equation}
The first term represents the usual hydrodynamic energy dissipation ($\eta$ being the viscosity of the
fluid) in the lubrication approximation, while the second term represents
the extra energy dissipation associated with the contact line motion over substrate. Here
$\xi_{\rm{cl}}$ is a phenomenological parameter representing the mobility of the contact line:
$\xi_{\rm{cl}}$ is infinitely large for pinned contact line, and is zero for freely moving contact line. Experiments~\cite{Li2013,Kajiya2009} showed that $\xi_{\rm{cl}}$ originates from the substrate wetting properties, substrate defects, and surface-active solutes. The first two factors affect the hydrodynamic dissipation at the contact line, while the third one effects the surface tension between liquid/vapor and liquid/substrate on the molecular scale, resulting different receding contact angle, then it further changes the mobility of the contact line ~\cite{Snoeijer2013}.

The velocity $v(r,t)$ in eq.~(\ref{f8}) is obtained from the solvent mass conservation equation
\begin{equation}\label{f9}
  \frac{d}{dt}\int^{r}_{0}dr' 2\pi r'h(r',t)
              = - 2\pi r v(r,t)h(r,t) - \int^{r}_{0}dr' 2\pi r' J(r',t)
\end{equation}
where $J(r, t)$ denotes the evaporation rate (the volume of solvent evaporating
per unit time per unit surface area) at position $r$ and time $t$.  It is known that
$J(r, t)$ diverges at the contact line in such a way
as  $J(r,t) \propto \left[R(t)-r\right]^{-\frac{1}{2}}$~\cite{Deegan1997}.  However
since this spatial dependence of $J(r,t)$ has a weaker effect compared with the effect of
contact line pinning, here we proceed assuming that $J(r, t)$ is independent of $r$ and
has the form
\begin{equation} \label{f10}
J(t)=-\frac{\dot{V}(t)}{\pi R^2(t)} = -\frac{\dot{V_0}}{\pi R_0R(t) }.
\end{equation}
Equations (\ref{f1}), (\ref{f9}) and (\ref{f10}) give the following
simple expression for $v(r,t)$
\begin{equation}\label{f11}
v(r,t)=r\left(\frac{\dot{R}}{2R}-\frac{\dot{H}}{4H}\right)
        =r \left(\frac{\dot{R}}{R}-\frac{\dot{V}}{4V}\right).
\end{equation}
Inserting this expression into eq. (\ref{f8}), the energy dissipation function $\Phi$ is calculated as
\begin{equation}\label{f12}
\Phi=\frac{3\pi^2\eta R^4}{4V} \left[\ln\left(\frac{R}{2\epsilon}\right)-1\right]
                                              \left(\dot{R}-\frac{R\dot{V}}{4V}\right)^2
       +\pi\xi_{\rm{cl}} R\dot{R}^2
\end{equation}
where $\epsilon$ is the molecular cut-off length which is introduced to remove the
divergence in the energy dissipation at the contact line.  Hereafter, we set $\epsilon=10^{-6}R_0$
for all calculations.

The Onsager principle states that
$\dot{R}$ is determined by the condition of $\partial (\Phi + \dot F) /\partial \dot R =0$.
By eqs.~(\ref{f7}) and (\ref{f12}), this gives the following evolution equation
\begin{equation}\label{f13}
\left(1+k_{\rm{cl}}\right)\dot{R}
         =  \frac{R\dot{V}}{4V}
             + \frac{\gamma_{LV}\theta\left(\theta^2-\theta^2_e\right)}{6 C \eta }
\end{equation}
where $C=\ln(R/2 \epsilon)-1$, and $k_{\rm{cl}}$ is defined by
\begin{equation}\label{f14}
   k_{\rm{cl}}=\frac{\xi_{\rm{cl}}\theta}{3C\eta}
\end{equation}
which characterizes the importance of the extra
friction constant $\xi_{\rm{cl}}$ of the contact line
relative to the normal hydrodynamic friction $\xi_{\rm{hydro}}=3C \eta/\theta$
\cite{Bonn09}. Since not much is known about $\xi_{\rm{cl}}$,
here we proceed making a simple assumption that the ratio
$k_{\rm{cl}}= \xi_{\rm{cl}}/\xi_{\rm{hydro}}$
is a constant, or a material parameter determined by droplet and substrate.

To simplify the equations, we define two time scales, the evaporation time $\tau_{\rm{ev}}$,
and the relaxation time $\tau_{\rm{re}}$
\begin{equation}\label{f15}
  \tau_{\rm{ev}}=\frac{V_0}{|\dot V_0|},
  \qquad
 \tau_{\rm{re}}=\frac{\eta V^{\frac{1}{3}}_0}{\gamma_{LV}\theta^{3}_{e}}
\end{equation}
The time $\tau_{\rm{ev}}$ represents the characteristic time for the droplet (of initial size $V_0$) to dry up,
and $\tau_{\rm{re}}$ represents the relaxation time, the time needed for the droplet
(initially having contact angle $\theta_0$) to have the equilibrium contact angle $\theta_e$.

By such definition, the evolution equation ( eq.~(\ref{f13}) ) becomes
\begin{equation}\label{f16}
  \left(1+k_{\rm{cl}}\right)\tau_{\rm{ev}} \dot{R}
     = -\frac{V_{0}R^2}{4R_0V}
             + \frac{V^{\frac{1}{3}}_{0}\theta \left(\theta^2-\theta^2_e\right)}
                            {6C k_{\rm{ev}} \theta^3_{e}
                             }
\end{equation}
where $k_{\rm{ev}}$ is defined by
\begin{equation}\label{f17}
k_{\rm{ev}}=\frac{\tau_{\rm{re}}}{\tau_{\rm{ev}}}
\end{equation}
which is another important parameter characterizing the drying behavior.
If $k_{\rm{ev}}$ is large, the droplet volume decreases much faster than the
equilibration of the contact angle $\theta$, therefore $\theta$ becomes much smaller than the
equilibrium value $\theta_e$.  On the other hand, if  $k_{\rm{ev}}$ is small,
$\theta$ remains close to  $\theta_e$.

For pure water or dilute polymer solutions of macroscopic size (diameter $1 mm$),
$k_{\rm{ev}}$ is less than $10^{-3}$.  On the other hand, for
concentrated polymer solutions with high viscosity $k_{\rm{ev}}$ can be larger than
$10^{-1}$~\cite{Kajiya09,Kajiya10}.

In eq.~(\ref{f16}), $V$ and $\theta$ are function of time. The evolution of $V(t)$ is given
by eq.~(\ref{f3}), which is written as
\begin{equation} \label{f18}
       \tau_{\rm{ev}}\dot{V}=-V_{0}\frac{R(t)}{R_0}.
\end{equation}
Finally, $\theta$ is related to $R$ and $V$ by
\begin{equation} \label{f19}
       \theta =\frac{4V}{\pi R^3}.
\end{equation}
Equations (\ref{f16}), (\ref{f18}) and (\ref{f19}) are the set of
equations which determine the time evolution of our system.

\begin{figure}[h]
\begin{center}
\includegraphics[bb=0 0 250 300, scale=1.0,draft=false]{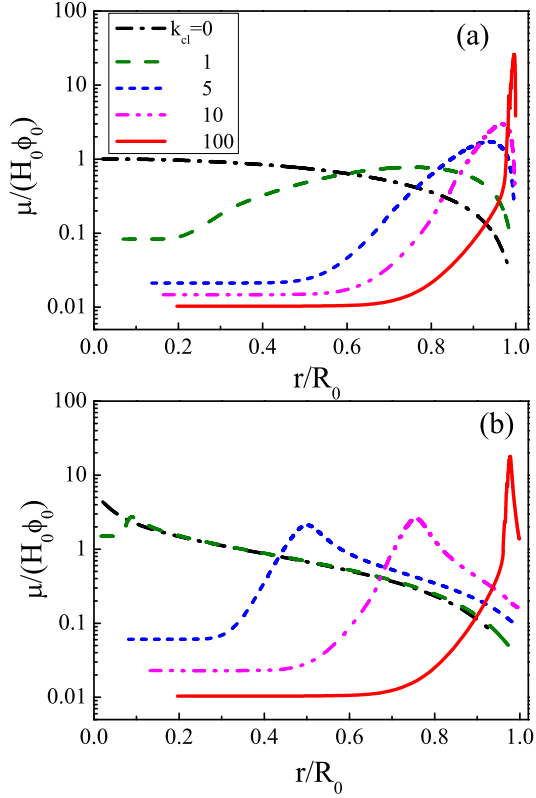}
\caption{(Color Online).
Profile of the deposit left on the substrate when the drying is completed.
Coffee-ring to volcano-like, then to mountain-like pattern transition induced by changing the value of $k_{\rm cl}$ from $100$ to $0$. (a) the case of fast evaporation rate characterized by $k_{\rm{ev}}=1$, and (b) the evaporation rate is small with $k_{\rm{ev}}=10^{-3}$. For both cases, $\theta_e=\theta_0=0.2$ and $\Delta t/\tau_{\rm{ev}}=10^{-5}$.}
\label{fig2}
\end{center}
\end{figure}

By using the same model, we can calculate the distribution of the deposit left on the substrate. We consider the solute located at position $r_0$ at time $t=0$. As solvent evaporates, such solute is convected by the fluid.  Let
$\tilde r(r_0,t)$ be the height-averaged position of such solute at time $t$.
Since the diffusion of the solute in radial direction can be ignored for
macroscopic droplets \cite{Anderson1995}, we can assume that the solute
moves with the same velocity as the fluid
as long as the solute is in the droplet, (i.e., as long as $\tilde r(r_0,t) <R(t)$ ),
\begin{equation}\label{f20}
    \dot {\tilde r}(r_0,t) =v\left[\tilde r(r_0,t),t\right]
                        = \left(\frac{\dot{R}}{R}-\frac{\dot{V}}{4V}\right)\tilde r(r_0,t)
                        \qquad \mbox{for} \quad \tilde r(r_0,t) < R(t)
\end{equation}
Such solute will be deposited on the substrate at the time $t_d$ which satisfies $R(t_d)= \tilde r(r_0,t_d)$
(Notice that $t_d$ defined in this way   is a function of $r_0$).
The total amount of solute which was originally contained in the region between $r_0$ and
$r_0+dr_0$ at time $t=0$ is $2 \pi r_0 h(r_0, 0) \phi_0 dr_0$, and this is
deposited in the region between $\tilde r $ and  $\tilde r + d \tilde r$.
Hence the density of deposit at the position
$\tilde r$ is given by
\begin{equation}\label{f21}
    \mu \left[\tilde r(r_0, t)\right] =h(r_0, 0) \phi_0 \frac{r_0}{\tilde r} \left( \frac{d \tilde r}{d r_0} \right)^{-1}.
\end{equation}
%

\begin{figure}[h]
\begin{center}
\includegraphics[bb=-130 0 530 251, scale=0.68,draft=false]{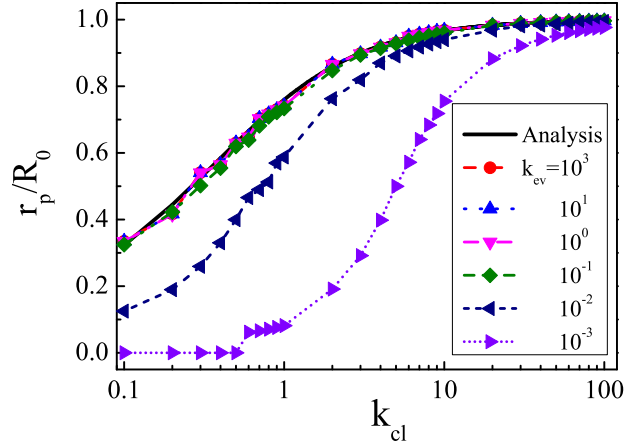}
\caption{(Color Online).
The peak position $r_p$ of the deposit density distribution is plotted against $k_{\rm{cl}}$.  The
results of numerical calculation for various values of $k_{\rm {ev}}$ are denoted by symbols,
while the analytical result, eq.(\ref{f26}), is denoted by the black-solid line.
All other parameters are the same as in Fig.\ref{fig2}.}
\label{fig3}
\end{center}
\end{figure}

Figure \ref{fig2} shows the deposition density obtained by such calculation.  Here the initial
contact angle $\theta_0$ and the equilibrium contact angle $\theta_e$ are
taken to be equal to 0.2.
Fig.~\ref{fig2}(a) corresponds to the case of fast evaporation where $k_{\rm{ev}}=1$, while Fig.~\ref{fig2}(b) corresponds to the case of slow evaporation where $k_{\rm{ev}}=10^{-3}$.
It is seen that in both cases the peak position shifts inward with the decrease of $k_{\rm{cl}}$.
When $k_{\rm{cl}}=100$, the contact line does not move much from the initial position $R_0$, and
the coffee-ring pattern appears.  On the other hand, when  $k_{\rm{cl}}=0 $,
most solute is accumulated at the center of the droplet, and a mountain-like pattern appears.

Figure~\ref{fig3} shows the peak position $r_p$ of the deposit profile $\mu(r)$
plotted as a function of $k_{\rm{cl}}$.  When $k_{\rm{cl}}$ is  very large,
$r_p$ is equal to $R_0$.  As $k_{\rm{cl}}$ decreases, $r_p$ decreases also.
For large values of $k_{\rm{ev}}$
($k_{\rm{ev}} \geq 0.1 $),  $r_p$ is independent of $k_{\rm{ev}}$, while
for small values of $k_{\rm{ev}}$ ($k_{\rm{ev}}<0.1$), $r_p$ becomes a function of $k_{\rm{ev}}$.

We can derive an analytical expression for $r_p$ for large values of $k_{\rm{ev}}$.
For  $k_{\rm{ev}} \gg 1$,  the second term in the right hand side of eq.~(\ref{f13}) can be
ignored, and the evolution equation for $R$ becomes
\begin{equation}\label{f22}
\dot{R}=\frac{1}{4(1+k_{\rm{cl}})}\frac{R\dot{V}}{V}.
\end{equation}
By using of eqs. (\ref{f11}) and (\ref{f22}), eq.~(\ref{f20}) is written as
\begin{equation}\label{f23}
 \dot{\tilde r}=-k_{\rm{cl}}\tilde r\frac{\dot{R}}{R}
\end{equation}
which is solved as $\tilde r(r_0,t) = r_0 \left[\frac{R(t)}{R_0}\right]^{-k_{\rm{cl}}}$.
The solute initially located at $r_0$ is deposited at time $t_d$ when $\tilde r(r_0,t_d)$ becomes
equal to $R(t_d)$, i.e., $r_0 \left[\frac{R(t_d) }{R_0}\right]^{-k_{\rm{cl}}}= R(t_d)$. This gives the following
relation between $\tilde r(r_0, t_d)= R(t_d)$ and $r_0$
\begin{equation}\label{f24}
    \tilde r(r_0, t_d)= R(t_d)= (r_0)^{\frac{1}{1+{k_{\rm{cl}}}}}(R_0)^{\frac{k_{\rm{cl}}}{1+k_{\rm{cl}}}}.
\end{equation}
Inserting eq.~(\ref{f24}) into the expression of $\mu$ in eq.~(\ref{f21}), we have
\begin{equation}\label{f25}
  \mu\left(\tilde r\right) =\phi_0 H_0 \left(1+k_{\rm{cl}}\right)
                                         \left(\frac{\tilde r}{R_0}\right)^{2k_{\rm{cl}}}
                                          \left[ 1 - \left( \frac{\tilde r}{R_0}\right )^{2(1+k_{\rm{cl}})}\right].
\end{equation}
By maximizing $\mu(\tilde r)$ with respect to $\tilde r$, we have an explicit expression for
the peak position $r_p$
\begin{equation}\label{f26}
    r_p=R_0 \left(\frac{k_{\rm{cl}}}{2k_{\rm{cl}}+1}\right)^{\frac{1}{2(1+k_{\rm{cl}})}}.
\end{equation}
This curve is shown by the black solid line in Fig.~\ref{fig3}, which agrees quite well with
numerical results. When $k_{\rm{cl}} \gg 1 $,  $r_p$ approaches to $R_0$,
giving the coffee-ring pattern (Deegan's limit)
, while when $k_{\rm{cl}} \ll 1 $, $r_p$ goes to $0$, leading to
the mountain-like deposition (Freed-Brown's limit).   Equation (\ref{f26}) smoothly
interpolates these two limits.

\begin{figure}[h]
\begin{center}
\includegraphics[bb=0 10 410 280, scale=0.88,draft=false]{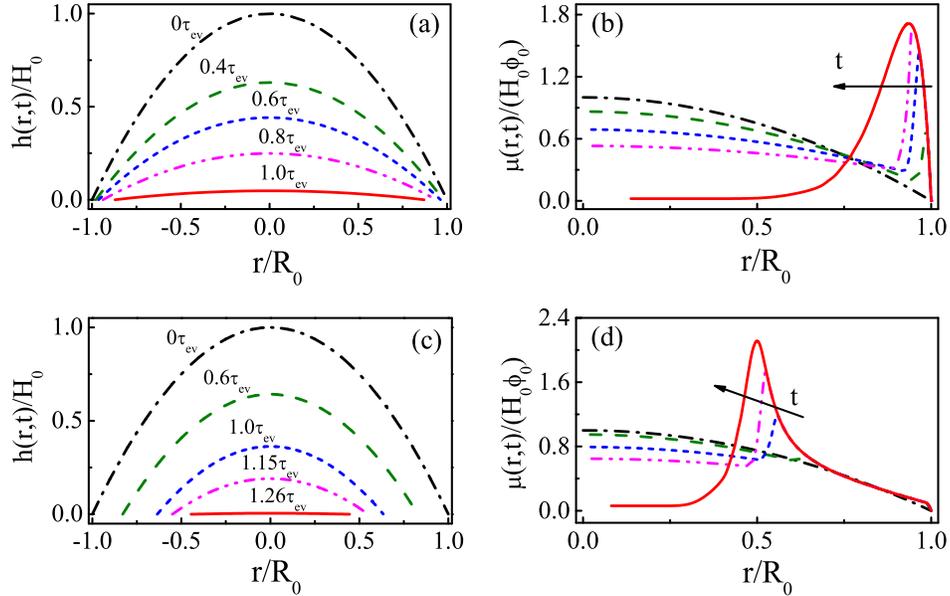}
\caption{(Color Online).
Evolution of the profile of droplet's interface between liquid and vapor, $h(r,t)/H_0$, and solute density distribution, $\mu(r,t)/(H_0\phi_0)$, for (a) and (b) fast evaporation rate, $k_{\rm{ev}}=1$, and for (c) and (d) slow evaporation rate, $k_{\rm{ev}}=10^{-3}$. In all figures, the black dash-dot line represents the initial state, while the red solid line is the final state after drying up. The time steps in (b) and (d) are the same as in (a) and (c) accrodingly. The arrow is the direction of time evolution of the drying prcess. Other parameters are $k_{\rm cl}=5$, $\theta_e=\theta_0=0.2$, and $\Delta t/\tau_{\rm{ev}}=10^{-5}$.}
\label{fig4}
\end{center}
\end{figure}

As it is seen in Fig.~\ref{fig3}, the curve $r_p/R_0$ vs $k_{\rm{cl}}$ starts to deviate from the
analytical formula (\ref{f26}) for $k_{\rm{ev}}\leq10^{-2}$. This deviation can be understood by looking at the droplet shape during drying.

Figure \ref{fig4} shows the time evolution of droplet shape and deposition pattern for both fast evaporation ($k_{\rm{ev}}=1$) and slow evaporation ($k_{\rm{ev}}=10^{-3}$). When the evaporation is fast, the solvent is taken out from the surface of the droplet inducing a large deformation of the droplet. Hence the contact angle $\theta(t)$ decreases in time, while the contact radius $R(t)$ remains almost constant (Fig.~\ref{fig4}(a)). As a result, the deposit peak appears near $R_0$ (Fig.~\ref{fig4}(b)).
On the other hand, when the evaporation rate is  slow, a fluid flow is induced to maintain the equilibrium contact angle.  Accordingly, the contact radius $R(t)$ decreases in time (Fig.~\ref{fig4}(c)), and the deposit peak appears inside the original contact area (Fig.~\ref{fig4}(d)).

Since there is no quantitative study about the relation between the drying condition and the
deposition pattern, it is difficult to make quantitative comparison between the
theory and experiments. However, qualitative comparison can be made. Li et al.~\cite{Li2013} studied various solution droplets placed on different substrates, and observed the transition of deposition pattern from coffee-ring to mountain-like. With weak contact angle hysteresis (CAH) substrate
(like silica glass or polycarbonate substrate for water droplet), the contact line recedes and forms the mountain-like deposition patterns. On the other hand, for strong CAH substrate (like graphite),
the contact line is pinned and leaves coffee-ring patterns. These observations are
in agreement with the present theory. Kajiya et al.~\cite{Kajiya2009} studied the drying of water-poly(N,N-dimethylacrylamide) PDMA droplet on glass substrate and observed volcano-like deposition patterns. They have explained the volcano-like pattern based on the mobility of the contact line,
which is also consistent with our theory.

In this letter, we have proposed a simple model of drying droplet which accounts for contact
line motion and solvent evaporation simultaneously. We have clarified how the contact line friction
(described by $k_{\rm{cl}}$) and the evaporation rate (described
by $k_{\rm{ev}}$) affect the final deposition pattern, especially the transition from
coffee-ring to volcano-like, then to mountain-like patterns. What remains to be done is to include a stick-slip mechanism in the model to capture the multi-ring phenomena and other more interesting deposition patterns.

This work was supported in part by grants 21404003 and 21434001 of the National Natural Science Foundation of China (NSFC), and the Fundamental Research Funds for the Central Universities.


\end{document}